\documentclass[journal]{IEEEtran}
\usepackage{amsmath,amsfonts}
\usepackage{algorithmic}
\usepackage{algorithm}
\pdfoutput=1
\usepackage{array}
\usepackage[caption=false,font=normalsize,labelfont=sf,textfont=sf]{subfig}
\usepackage{textcomp}
\usepackage{stfloats}
\usepackage{url}
\usepackage{verbatim}
\usepackage{graphicx}
\usepackage{cite}
\usepackage{xcolor}
\usepackage{bm}
\usepackage{amsmath,amssymb,amsfonts}

\begin{document}

\title{{\LARGE Double-RIS Aided
Multi-user MIMO Communications: Common Reflection Pattern and Joint Beamforming Design}}

\author{Pingping Zhang, Shiqi Gong, and Shaodan Ma,~\IEEEmembership{Senior Member,~IEEE}
\thanks{Pingping Zhang and Shaodan Ma are with the State Key Laboratory of Internet of Things for Smart City and the Department of Electrical and Computer Engineering, University of Macau, Macau, China (e-mail: yc17433@um.edu.mo; shaodanma@um.edu.mo).}
\thanks{S. Gong is with the School of Cyberspace Science and Technology, Beijing Institute of Technology, Beijing 100081, China (e-mail: shiqigong@bit.edu.cn).}}



\maketitle
\begin{abstract}
    — 
    Reconfigurable intelligent
surface (RIS) has entered the public consciousness
 as a promising technology for enhancing the performance of future wireless communication systems by dynamically constructing the wireless channels.
In this letter, we study a double-RIS aided downlink multi-user multiple-input multiple-output (MIMO) communication system. We investigate the mean-square-error (MSE) minimization problem by jointly optimizing the active transmit beamforming, the receive equalizer and the passive beamforming at each RIS. Different from prior works, for the sake  of reducing both communication overhead  and signal processing complexity, we assume that the two RISs utilize the common reflection pattern. Under this assumption, the coupling of the variables becomes tighter, thereby making the optimization problem more challenging to solve. To effectively address this issue, we propose a majorization-minimization (MM)-based alternating optimization (AO) algorithm.
 Numerical results show that in high signal-to-noise ratio (SNR) region, the double-RIS with common reflection pattern can achieve nearly the same performance as 
that with separate reflection pattern whereas the  complexity is only half of the latter. Thus, our proposed design enables an effective tradeoff between the  performance  and  the  implementation complexity of the considered system.
\end{abstract}

\begin{IEEEkeywords}
    Double-RIS, multi-user MIMO system, MSE, common reflection pattern.
\end{IEEEkeywords}

\vspace{-3mm}

\section{Introduction}
\IEEEPARstart{A}{s} a key enabling technology for the future generation wireless systems, reconfigurable intelligent
surface (RIS) can adaptively reconfigure the wireless communication environment to improve both energy and spectrum  efficiency\cite{9429987,9925080}. 
In general, RIS is an intelligent metasurface made up of a large number of passive reflecting elements. Each element can independently adjust the input  electromagnetic (EM) signal's amplitude and phase\cite{wu2021intelligent}.
In contrast to traditional multiple-input multiple-output (MIMO) relay communications \cite{gong2016millimeter}, RIS can provide higher beamforming gains while consuming less power.
Moreover, thanks to additional advantages of easy deployment, environment friendly, high compatibility and low cost\cite{liu2021reconfigurable}, RIS has been widely used in wireless communication systems.

Most existing works only considered the passive beamforming design for the single-RIS scenario \cite{zhang2020capacity,9739078,9279253}. For example,\cite{zhang2020capacity} jointly designed the transmit covariance matrix and RIS reflection pattern matrix by investigating the channel capacity maximization problem considering a single-user MIMO communication system aided by one RIS.
The work in \cite{9739078} provided a general framework for the transceiver designs in
the single-RIS aided 
 single-user and multi-user MIMO communication systems.
To further explore the RIS's potential in enhancing wireless communication performance, some works extend the single-RIS scenario to the multi-RIS scenario. For example,
\cite{xu2022statistically} studied the  mean-square-error (MSE) minimization problem by jointly optimizing the RIS phase shifts, the transmit beamforming and the receive equalizers  considering a multiple but non-cooperative RISs aided MIMO system.
The work in \cite{9714463} aimed to maximize  the channel capacity considering a cooperative double-RIS empowered single-user MIMO system  with line-of-sight (LoS) channel, while
 \cite{zheng2021double} further extended the above study to the cooperative double-RIS assisted multi-user MIMO system. The authors jointly optimized the receive beamforming  and the cooperative reflection coefficients at two RISs aiming to maximize the worst signal-to-interference-plus-noise ratio (SINR) of all users.

In this letter, from  the perspective  of reducing the high  communication overhead and signal processing complexity  induced by the large-scale RIS, we consider the cooperative double-RIS aided downlink multi-user MIMO communication system with the common reflection
pattern.
 We resort to minimize the  average sum MSE of all data symbols by jointly optimizing the active transmit  beamforming matrices, the receive equalizer  and the common RIS reflection pattern. The formulated optimization problem is generally more challenging to solve than its counterpart with separate reflection pattern due to the strongly coupled optimization variables. To tackle this difficult problem,  a
majorization-minimization (MM)-based alternating optimization (AO) algorithm is proposed.
Numerical simulation results  illustrate the better performance of the proposed algorithm when compared with the single-RIS system, and also show that our proposed algorithm can perform as best as the double-RIS system with the separate reflection pattern in the high-SNR region. 

\vspace{-2.3mm}

\section{  Problem Formulation And System Model}
As shown in Fig. 1, we consider a double-RIS aided downlink MIMO communication system with ${K}$ users, where each user is equipped with ${N}_{r,k}$ antennas and the BS is equipped with ${N}_{t}$ antennas.
The two distributed RISs, namely, RIS 1 and RIS 2, are assumed to be near the BS and the users, respectively, so that the productive path loss of the double-RIS cascaded channel is minimized\cite{9714463}. Since the two RISs utilize the common reflection pattern, we assume they have the equal number of reflection elements, denoted as $M$. The direct channel between the BS and the users is ignored due to the blockages.

Let $\mathbf {T}_{l}\in \mathbb {C}^{M\times N_{t}}$, $\mathbf {S}\in \mathbb {C}^{M\times M}$ and $\mathbf {R}_{l,k}\in \mathbb {C}^{N_{r,k}\times M}$ denote the channel matrices for the  BS $\rightarrow$ RIS ${l}$, RIS ${1\rightarrow}$  RIS ${2}$ and RIS ${l\rightarrow}$ user ${k}$ links, respectively, with $l=1,2$ and $k=1,2,...,K$.
The common reflection coefficients of two RISs are introduced as ${ \boldsymbol \Theta } =  \mathop {\mathrm {diag}}\nolimits \{\theta  _{1},\cdots,\theta  _{M}\}$, where the amplitude of each element is assumed to be 1.
Under the aforementioned settings, an effective cascaded channel between the BS and the ${k}$th user is expressed as
\begin{equation} \mathbf {H}_{k} =\mathbf {R}_{1,k} \boldsymbol {\Theta }\mathbf {T}_{1}+
  \mathbf {R}_{2,k} \boldsymbol {\Theta  }\mathbf {T}_{2}+
  \mathbf {R}_{2,k} \boldsymbol {\Theta }\mathbf {S} \boldsymbol {\Theta  }\mathbf {T}_{1}. 
  \label{channel}\end{equation}

We assume total ${N}$ data streams are transmitted from the BS to all ${K}$ users and ${\mathbf{x}} \triangleq \left [{ {\mathbf{x}}_{1}^{\mathrm {H}}, \cdots, {\mathbf{x}}_{K}^{\mathrm {H}}}\right]^{\mathrm {H}}\in \mathbb {C}^{N \times 1}$ is defined as the transmitted data  symbol vector, where ${\mathbf{x}}_{k} \in \mathbb {C}^{N_{k} \times 1}$ is the data symbols  for the $k$th user, obeying Gaussian distribution, i.e., $\mathbf{x}_{k} \sim \mathcal {CN}(\mathbf{0}, {\mathbf{I}}_{N_{k}})$ and  $\sum _{k = 1}^{K} {N_{k}} = N$.  ${\mathbf{F}} \triangleq \left [{ {\mathbf{F}}_{1}, \cdots, {\mathbf{F}}_{K} }\right]  \in \mathbb {C}^{N_{t} \times N}$ is defined as the transmit beamformer, where ${\mathbf{F}}_{k} \in \mathbb {C}^{ N_{t} \times N_{k}}$ is  the $k$th user's beamformer. For simplicity, we assume that the BS has perfect knowledge of system global  channel state information (CSI).  The received signal ${\mathbf{y}}_{k} \in \mathbb {C}^{N_{r,k} \times 1}$ at the $k$th user side is then given by
\begin{equation} \mathbf {y}_{k} = \mathbf {H}_{k}\mathbf {F}\mathbf {x}+\mathbf {z}_{k}, \end{equation}
where $\mathbf{z}_{k} \sim \mathcal {CN}(\mathbf{0},\sigma ^{2} {\mathbf{I}}_{N_{r,k}})$ is the additive white Gaussian noise (AWGN).  For the $k$th user, upon receiving ${\mathbf{y}_{k}}$, the equalizer ${\mathbf{G}_{k}} \in \mathbb {C}^{{N}_{k} \times N_{r,k}}$ is used to decode the transmitted signal, i.e.,
$\hat {\mathbf {x}}_{k}={\mathbf{G}_{k}}\mathbf {y}_{k}$.
Accordingly, the MSE matrix of the estimated signal $\hat {\mathbf {x}}_{k}$ is expressed as
\begin{equation}\begin{aligned}
& \boldsymbol {\Psi }_{\mathrm{ MSE,k}} ({\mathbf{F}_{k}}, {\mathbf{G}_{k}}, \boldsymbol {\Theta }) \\ &=\mathbb {E} \left \lbrace{ ( {\mathbf {x}}_{k} - \hat{\mathbf{x}_{k}})^{\mathrm {H}} ( {\mathbf {x}}_{k} - \hat{\mathbf{x}_{k}})}\right \rbrace \\&= {\mathbf{G}}_{k} {\mathbf{H}}_{k} {\mathbf{F}}( {\mathbf{G}}_{k} {\mathbf{H}}_{k} {\mathbf{F}}) ^{\mathrm {H}}\!\! -\!\,2{\Re}( {\mathbf{G}}_{k} {\mathbf{H}}_{k} {\mathbf{F}}_{k})  \!   + \!\sigma _{n}^{2} {\mathbf{G}}_{k} {\mathbf{G}}_{k}^{\mathrm {H}}\!\!+\!\!{\mathbf{I}}_{{N}_{k}}.  \end{aligned}\end{equation}

In this letter, we aim to minimize the average sum MSE of all data symbols  by jointly designing the BS transmit beamforming matrices $\{ {\mathbf{F}}_{k} \}_{k=1}^{K}$, the linear receive equalizers $\{ {\mathbf{G}}_{k} \}_{k=1}^{K}$ and
the common RIS reflection pattern $\boldsymbol {\Theta }$, subject to the total BS transmit power constraint and the unit-modulus constraints for all RIS reflecting elements. The corresponding optimization problem is formulated as 

\vspace{-5mm}
\begin{subequations}\begin{align}\text {(P1)} &\quad \min _{\boldsymbol{ \{ {\mathbf{F}}_{k} , {\mathbf{G}}_{k} \}_{k=1}^{K},{\Theta } }}~ \sum _{k = 1}^{K}{\mathrm {tr}} ({\boldsymbol {\Psi }_{\mathrm{ MSE,k}} ({\mathbf{F}_{k}}, {\mathbf{G}_{k}},\boldsymbol {\Theta  }) } )\label{5a}
\\&\qquad \quad ~\text {s.t.}~
     {\mathrm {tr}} ( {\mathbf{F}} {\mathbf{F}}^{\mathrm {H}})\le P, \label{5b}
     \\&\hphantom {\qquad \quad ~\text {s.t.}~} 
       \boldsymbol {\Theta }= \mathop {\mathrm {diag}}\nolimits \{\theta  _{1},\cdots,\theta  _{M}\},\label{5c}
       \\&\hphantom {\qquad \quad ~\text {s.t.}~} |\theta  _{{i}}|=1, \quad i\in \{1,\cdots,M\},\label{5d}
     \end{align}  \end{subequations}    
where ${P}$ is the BS maximum transmit power.
\begin{figure}[htbp]
    \centering
    \includegraphics[width=3.1in]{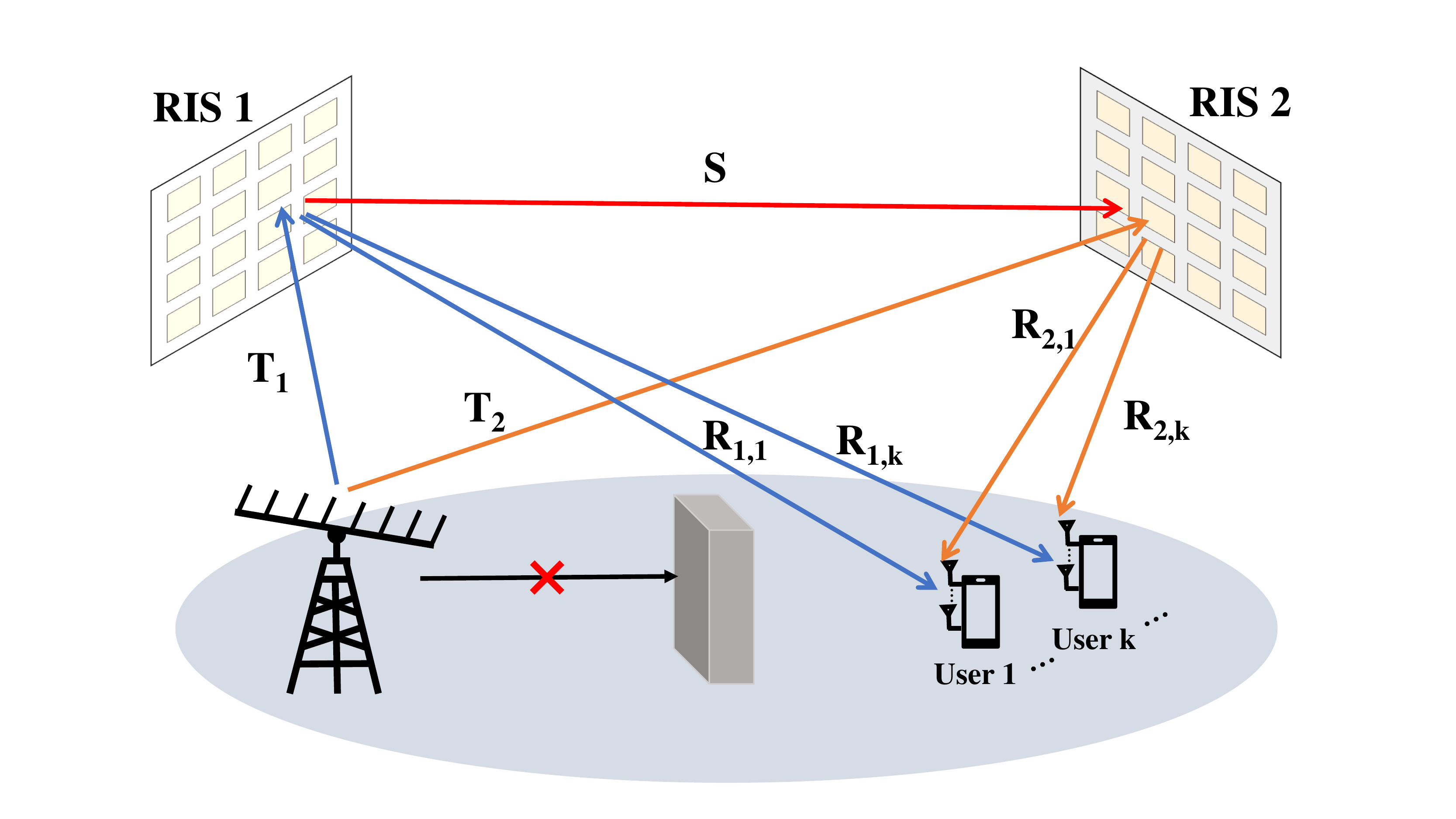}
    \caption{A double-RIS aided multi-user MIMO system.}
    \label{fig_1}
    \end{figure}
It can be observed that problem (P1) is challenging to solve due to
the strongly coupled variables in the objective function (\ref{5a}) and
the non-convex  constraints in (\ref{5d}).
Moreover, the adopted common RIS reflection pattern leads to an intractable fourth-order term in the objective function (\ref{5a}), which also greatly increases the difficulty of solving problem (P1) when compared to the separate reflection pattern case. Next, we present an MM-based AO algorithm 
to decouple the variables and tackle the intractable high-order term caused by the common reflection pattern.

\vspace{-2.5mm}

\section{Proposed MM-Based AO Algorithm }
In this section,  an MM-based AO algorithm is proposed to solve the challenging problem (P1), where each optimization variable is updated while keeping the others fixed. By leveraging the MM technique, each subproblem admits a closed-form optimal solution. 

\vspace{-3mm}

\subsection{ Equalizer and Transmit Beamforming Optimization}
Firstly, we derive the optimal equalizer $\mathbf {G}_{k}$ with the fixed $\{ {\mathbf{F}}_{k} \}_{k=1}^{K}$ and $\mathbf {\Theta}$. In this case, It can be easily found that the problem (P1) w.r.t $\mathbf {G}_{k}$ is an unconstrained convex problem. Based on the LMMSE criterion, the optimal ${\mathbf{G}}_{k}^{\mathrm{ opt}} $ can be derived as
\begin{equation} {\mathbf{G}}_{k}^{\mathrm{ opt}} = {\mathbf{F}_{k}}^{\mathrm {H}}{\mathbf{H}_{k}}^{\mathrm {H}}({\mathbf{H}_{k}} {\mathbf{F}} {\mathbf{F}}^{\mathrm {H}}{\mathbf{H}_{k}}^{\mathrm {H}}+
     {\mathbf{R}}_{\mathbf{z}_{k}})^{-1}.\label{6}\end{equation}

Secondly, with the fixed $\{ {\mathbf{G}}_{k}\}_{k=1}^{K}$ and $\mathbf {\Theta}$, we optimize the  active transmit beamforming $\mathbf {F}_{k}$. We reformulate problem (P1)  as 

\vspace{-7mm}
\begin{subequations}\begin{align}&\text {(P2)} \quad \min _{\mathbf{ F}_{k}}~ \sum _{k = 1}^{K}{\mathrm {tr}} (\boldsymbol {\Psi }_{\mathrm{ MSE,k}} ({\mathbf{F}_{k}})),
    \\&\qquad \quad ~\text {s.t.}~{\mathrm {tr}} ( {\mathbf{F}} {\mathbf{F}}^{\mathrm {H}})\le P.
    \end{align} \end{subequations} 
 Problem (P2) can be easily found to be a quadratic convex problem
 w.r.t. $\mathbf {F}_{k}$. So the Karush-Kuhn-Tucker (KKT) conditions can be utilized to obtain the optimal $\mathbf {F}_{k}$. Specifically, defining $\mu$ as the Lagrange multiplier, problem (P2)'s  Lagrangian function is expressed as
\begin{equation} L ({{\mathbf{F}_{k}},\;\mu }) =\sum _{k = 1}^{K}{\mathrm {tr} (\boldsymbol {\Psi }_{\mathrm{ MSE,k}} ({\mathbf{F}_{k}})) }+ \mu (\mathrm {tr}\left \lbrace{ {\mathbf{F}} 
    {\mathbf{F}}^{\mathrm {H}}}\right \rbrace - P).\end{equation}
Then the KKT conditions can be derived as
\begin{subequations}\begin{align}&\mathbf {F}_{k} = (\sum _{i = 1}^{K}{\mathbf {H}_{i}^{H} \mathbf {G}_{i}^{H} \mathbf {G}_{i} \mathbf {H}_{i} + \mu \mathbf {I}_{Nt} })^{-1}  \mathbf {H}_{k}^{\mathrm{H}}\mathbf {G}_{k}^{\mathrm{H}},\label{9a}
    \\[2pt]&\mu ({\mathrm {tr} \lbrace{ {\mathbf{F}} {\mathbf{F}}^{\mathrm {H}}} \rbrace - P }) = 0,\;\mu \ge 0,\label{9b}
    \\[2pt]&\mathrm {tr} \lbrace{ {\mathbf{F}} {\mathbf{F}}^{\mathrm {H}}} \rbrace - P \le 0.\end{align}\end{subequations}

It follows from (\ref{9a}) that the optimal $\mathbf {F}_{k}$ depends on the Lagrange multiplier $\mu$. Specifically, if  ${\mathrm {tr}} ( {\mathbf{F}} {\mathbf{F}}^{\mathrm {H}}) 
\le P$ and $\mu=0$, we have $\mathbf {F}_{k} = (\sum _{i = 1}^{K}{\mathbf {H}_{i}^{H} \mathbf {G}_{i}^{H} \mathbf {G}_{i} \mathbf {H}_{i}} )^{-1}  \mathbf {H}_{k}^{\mathrm{H}}\mathbf {G}_{k}^{\mathrm{H}}$.
If $\mu>0$, we have $\mathrm {tr} \lbrace{ {\mathbf{F}} {\mathbf{F}}^{\mathrm {H}}} \rbrace = P$.
Let $\sum _{i = 1}^{K}{\mathbf {H}_{i}^{H} \mathbf {G}_{i}^{H} \mathbf {G}_{i} \mathbf {H}_{i}}=\mathbf {U}{\Lambda }\mathbf {U}^{H}$, where $\boldsymbol{\Lambda }=\mathop {\mathrm {diag}}\nolimits \{\lambda_{1},\ldots,\lambda_{M}\}$ is the diagonal matrix consisting of eigenvalues  of
   $\sum _{i = 1}^{K}{\mathbf {H}_{i}^{H} \mathbf {G}_{i}^{H} \mathbf {G}_{i} \mathbf {H}_{i}}$.
   Substituting  this decomposition into (\ref{9a}), the BS transmit power is monotonically decreasing over $\mu$ as follows:

\vspace{-4mm}
   
  \begin{equation} \begin{aligned} {\mathrm {tr}} (\mathbf {F}\mathbf {F}^{H})
   =&{\mathrm {tr}} [\sum _{k = 1}^{K} (\sum _{i = 1}^{K}{\mathbf {H}_{i}^{H} \mathbf {G}_{i}^{H} \mathbf {G}_{i} \mathbf {H}_{i} + \mu \mathbf {I}_{Nt} })^{-2} { \mathbf {H}_{k}^{H}\mathbf {G}_{k}^{H}\mathbf {G}_{k} \mathbf {H}_{k} }] \\
       =&{\mathrm {tr}}  [{  ({\boldsymbol{\Lambda } + {\mu } \mathbf {I}_{N} })^{-2} \mathbf {R} }] = \sum _{i=1}^{N} \frac {[\mathbf {R}]_{i,i}}{(\lambda _{i} + {\mu })^{-2}},\label{10}\end{aligned}\end{equation}
   where $\mathbf {R}\triangleq\mathbf {U}^{H}(\sum _{k = 1}^{K}{ \mathbf {H}_{k}^{H}\mathbf {G}_{k}^{H}\mathbf {G}_{k} \mathbf {H}_{k} })\mathbf {U}$.
   Thus, the optimal  $\mu$ can be efficiently found through one-dimensional search methods, e.g., bisection search\cite{shi2011iteratively}. Then the optimal $\mathbf {F}_{k}$ can be obtained from (\ref{9a}).

\vspace{-3mm}
   
\subsection{ Common Reflection Pattern Optimization}


After obtaining $\{ {\mathbf{F}}_{k} \}_{k=1}^{K}$ and $\{ {\mathbf{G}}_{k} \}_{k=1}^{K}$, we aim to optimize the common RIS reflection pattern $\mathbf {\Theta}$. Different from the traditional double-RIS system with the separate reflection pattern, the common reflection pattern makes the objective function more complicated and higher-order. Instead of directly solving it, we explore its inherent structure to simplify the problem. To be specific, by omitting the irrelevant constants, the objective function can be written as 

\vspace{-5mm}

\begin{equation}\begin{aligned}  & f_{\mathrm{ Obj}} ( \boldsymbol {\Theta })
     \\=&\sum _{k = 1}^{K}[\mathrm{tr}(\sum _{i = 1}^{2}\sum _{j = 1}^{2}{\mathbf{G}_{k}}\mathbf {R}_{i,k} \boldsymbol {\Theta }\mathbf {T}_{i} {\mathbf{F}}{\mathbf{F}}^{\mathrm {H}}\mathbf {T}_{j}^{\mathrm{H}} \boldsymbol {\Theta }^{\mathrm {H}}\mathbf {R}_{j,k}^{\mathrm{H}}\mathbf {G}_{k}^{\mathrm{H}}
     \\&+\sum _{i = 1}^{2}({\mathbf{G}_{k}}\mathbf {R}_{i,k} \boldsymbol {\Theta }\mathbf {T}_{i} {\mathbf{F}}{\mathbf{F}}^{\mathrm {H}}\mathbf {T}_{1}^{\mathrm{H}} \boldsymbol {\Theta }^{\mathrm {H}}\mathbf{S}^{\mathrm{H}} \boldsymbol {\Theta }^{\mathrm {H}}\mathbf {R}_{2,k}^{\mathrm{H}}\mathbf {G}_{k}^{\mathrm{H}}
      \\&+{\mathbf{G}_{k}}\mathbf {R}_{2,k} \boldsymbol {\Theta }\mathbf{S} \boldsymbol {\Theta }\mathbf {T}_{1} {\mathbf{F}}{\mathbf{F}}^{\mathrm {H}}\mathbf {T}_{i}^{\mathrm{H}} \boldsymbol {\Theta }^{\mathrm {H}} \mathbf {R}_{i,k}^{\mathrm{H}}\mathbf {G}_{k}^{\mathrm{H}})
    \\&+{\mathbf{G}_{k}}\mathbf {R}_{2,k} \boldsymbol {\Theta }\mathbf{S} \boldsymbol {\Theta }\mathbf {T}_{1} {\mathbf{F}}{\mathbf{F}}^{\mathrm {H}}\mathbf {T}_{1}^{\mathrm{H}} \boldsymbol {\Theta }^{\mathrm {H}}\mathbf{S}^{\mathrm{H}} \boldsymbol {\Theta }^{\mathrm {H}}\mathbf {R}_{2,k}^{\mathrm{H}}\mathbf {G}_{k}^{\mathrm{H}})
    \\&-2{\Re}(\mathrm {tr}(\sum _{i = 1}^{2}\mathbf{G}_{k}\mathbf {R}_{i,k} \boldsymbol {\Theta }\mathbf {T}_{i} {\mathbf{F}}\!+\!{\mathbf{G}_{k}}\mathbf {R}_{2,k} \boldsymbol {\Theta }\mathbf{S} \boldsymbol {\Theta }\mathbf {T}_{1}{\mathbf{F}}))].
    \end{aligned}\end{equation}

\vspace{-2.8mm}
    
By  stacking   the diagonal elements of  RIS reflection coefficients $\mathbf {\Theta}$ and a general matrix $\mathbf {P}$  into  the vectors $\boldsymbol {\theta } \triangleq {\mathrm {diag}(\mathbf {\Theta})}$ and $\mathbf {p } \triangleq{\mathrm {diag}(\mathbf {P})}$, respectively, 
and using the identities
 $\mathrm {tr}(\bf {X} \bf {Y} \bf {Z} \bf {W})=\mathrm {vec}(\bf {X}^{ \mathrm {H} })^{ \mathrm {H} }(\bf {W}^{\mathrm {T}} \otimes \bf {Y}) \mathrm {vec}(\bf {Z})$ and $\mathrm {tr}(\boldsymbol {\Theta }\mathbf{P})=\boldsymbol {\theta }^{\mathrm{H}}{{\mathbf{p}}^{\ast}}$, 
we have

\vspace{-4mm}

\begin{equation}\begin{aligned} &f_{\mathrm{ Obj}} ( \boldsymbol {\Theta }) 
     \\&  \!\! =\! \!\mathrm {vec}(\!\boldsymbol {\Theta }\!)^{\mathrm{\!H}}\!\mathbf{A}\!\mathrm {vec}(\!\boldsymbol {\Theta }\!)\!\!+\!\!\mathrm {vec}(\!\boldsymbol {\Theta }\!)^{\mathrm{H}}\mathbf{B}\mathrm {vec}(\!\boldsymbol {\Theta }\mathbf{S} \boldsymbol {\Theta }\!)
        \!\!+\!\!\mathrm {vec}(\!\boldsymbol {\Theta }\mathbf{S} \boldsymbol {\Theta }\!)^{\mathrm{H}}\mathbf{B}^{\mathrm{H}}\mathrm {vec}(\!\boldsymbol {\Theta }\!)
       \\& \ +\!\!\mathrm {vec}(\!\boldsymbol {\Theta }\mathbf{S} \boldsymbol {\Theta }\!)^{\mathrm{H}}\mathbf{C}\mathrm {vec}(\!\boldsymbol {\Theta }\mathbf{S} \boldsymbol {\Theta }\!)
    \!  \! -\!\!2{\Re}(\boldsymbol {\theta }^{\mathrm{H}}{{\mathbf{p}}^{\ast}}\!\!+\!\!\mathrm {vec}(\!\boldsymbol {\Theta }^{\mathrm{H}}\!)^{\mathrm{H}}\mathbf{D}\mathrm {vec}(\!\boldsymbol {\Theta }\!)\!),
  \label{11}  \end{aligned}\end{equation}
where 

\vspace{-4mm}

     \begin{equation}\begin{aligned}
     \mathbf{A}=&\sum _{i = 1}^{2}\sum _{j = 1}^{2}((\mathbf {T}_{i} {\mathbf{F}}{\mathbf{F}}^{\mathrm {H}}\mathbf {T}_{j}^{\mathrm{H}})^{\mathrm {T}} \otimes\sum _{k = 1}^{K}\mathbf {R}_{j,k}^{\mathrm{H}}\mathbf {G}_{k}^{\mathrm{H}}\mathbf{G}_{k}\mathbf {R}_{i,k}),
 \\\mathbf{B}=&\sum _{i = 1}^{2}(\mathbf {T}_{1} {\mathbf{F}}{\mathbf{F}}^{\mathrm {H}}\mathbf {T}_{i}^{\mathrm{H}})^{\mathrm {T}} \otimes\sum _{k = 1}^{K}\mathbf {R}_{i,k}^{\mathrm{H}}\mathbf {G}_{k}^{\mathrm{H}}\mathbf{G}_{k}\mathbf {R}_{2,k}, 
 \\\mathbf{C}=&(\mathbf {T}_{1} {\mathbf{F}}{\mathbf{F}}^{\mathrm {H}}\mathbf {T}_{1}^{\mathrm{H}})^{\mathrm {T}} \otimes\sum _{k = 1}^{K}\mathbf {R}_{2,k}^{\mathrm{H}}\mathbf {G}_{k}^{\mathrm{H}}\mathbf{G}_{k}\mathbf {R}_{2,k},
 \\\mathbf{D}=& \sum _{k = 1}^{K}(\mathbf {T}_{1}\mathbf {F}\mathbf {G}_{k}\mathbf {R}_{2,k})^{\mathrm{\!T}}\otimes\mathbf{S}, 
 \mathbf{P}=\sum _{k = 1}^{K}(\sum _{i = 1}^{2}\mathbf {T}_{i} {\mathbf{F}_{k}}\mathbf{G}_{k}\mathbf {R}_{i,k}).
 \end{aligned}\end{equation}
 
 Thanks to the sparse structure of $\mathrm {vec}(\boldsymbol {\Theta })$, the objective function ({\ref{11}) can be further simplified. Specifically, 
  let $\mathbf {A}_{0}\!\in \!\mathbb {C}^{M \times M}$ and $\mathbf{D}_{0} \!\in \!\mathbb {C}^{M \times M}$  be made up of all elements at the intersections of the $[(m\!- \!1) M \!+\!m {\;\;\!\!\!}]$th column and the $[(n - 1) M + n {\;\;\!\!\!}] $th row of 
  $\mathbf {A}$ and $\mathbf {D}$ for $m, n\! \!=\!\! 1,\! \ldots \!{}, M$, respectively. $\mathbf{B}_{0}$ is made up of all elements of the columns $[(q\! - \!1) M + q {\;\;\!\!\!}], q\! \!= \!\!1, \ldots {}, M $ of $\mathbf {B}$ and $\mathbf {C}_{0}\!\!=\!\!\mathbf {C}$. We define $\mathbf {v}\!\triangleq\!\mathrm {vec}(\boldsymbol {\Theta }\mathbf{S} \boldsymbol {\Theta })\!\!=\!\!(\boldsymbol {\Theta}^{\mathrm{T}}\!\otimes\!\boldsymbol {\Theta})\mathrm {vec}(\mathbf{S})$.
  Then the subproblem in terms of the common RIS reflection pattern is expressed as
 \begin{equation} \begin{aligned} \!\!\text {(P3)\!\!\!} \!\quad &\underset { \boldsymbol {\theta}}{\min }~ \!\boldsymbol{\theta}^{\mathrm{\!H}}\!\mathbf{A}_{0} \boldsymbol {\theta} \!\!+\!\! \boldsymbol {\theta}^{\mathrm{\!H}}\mathbf{B}_{0}\mathbf {v} \!\! +\!\! \mathbf {v}^{\mathrm{\!H}} {\mathbf {B}} ^{\mathrm{\!H}}_{0}\boldsymbol {\theta} 
     \!\! + \!\!\mathbf {v}^{\mathrm{\!H}} \mathbf {C}_{0} \mathbf {v} 
{\!-\!2{\Re}(\boldsymbol {\theta }^{\mathrm{\!H}}\mathbf{p}^{\!\ast}\!\!+\!\!\boldsymbol {\theta}^{\mathrm{\!H}} \mathbf {D}_{0}^{\!\ast} \boldsymbol {\theta}^{\!\ast})}
      \\ & ~\text {s.t.}~
       \mathbf {v}=(\boldsymbol {\Theta}^{\mathrm{T}}\otimes\boldsymbol {\Theta})\mathrm {vec}(\mathbf{S}),
      \\ &\hphantom { ~\text {s.t.}~}\mathbf {\Theta }= \mathop {\mathrm {diag}}\nolimits \{\theta  _{1},\cdots,\theta  _{M}\},
      \\ &\hphantom { ~\text {s.t.}~} |\theta  _{i}|=1, \quad i\in \{1,\cdots,M\} .
       \end{aligned}\end{equation}

 Problem (P3) is still hard to get the solutions due to the coupling of  $\boldsymbol {\Theta}$ and  $\mathbf{S}$ in the vector $\mathbf {v}$. Next, we resort to separate them.
Define $\mathbf {B}_{0}\triangleq[\mathbf {B}_{0,1}, \ldots {}, \mathbf {B}_{0,M}]\in \mathbb {C}^{M \times M^{2}}$, where $\mathbf {B}_{0,p}\in \mathbb {C}^{M \times M}$ is the block matrix consisting of the $[(p - 1) M + 1 {\;\;\!\!\!}]$th to the $[pM]$th column of $\mathbf {B}_{0}$ for $p\! \!=\!\! 1,\! \ldots \!{}, M$. $\mathbf {C}_{0}$ is similarly defined as $\mathbf {C}_{0}\triangleq\begin{bmatrix}
\mathbf{C}_{0,1,1}& \!\ldots\! {}&\mathbf {C}_{0,1,M}\\
\!\vdots\! {}&\ddots&\vdots {}\\
\mathbf {C}_{0,M,1}& \ldots {}&\mathbf {C}_{0,M,M}
\end{bmatrix}$,
where $\mathbf {C}_{0,j,l}\in \mathbb {C}^{M \times M}$ is the block matrix consisting of the $[(j - 1) M + 1 {\;\;\!\!\!}]$th row to the $[jM]$th row and the $[(l - 1) M + 1 {\;\;\!\!\!}]$th column to the $[lM]$th column of $\mathbf {C}_{0}$ for $j,l\! \!=\!\! 1,\! \ldots \!{}, M$. Then we have

\vspace{-5mm}

\begin{equation} \begin{aligned} 
{\boldsymbol {\theta}}^{\mathrm{H}}\mathbf {B}_{0}\mathbf {v}
=&\sum _{i=1}^{M}\boldsymbol{\theta}^{\mathrm{H}}{\mathbf {B}}_{0,i}{\mathrm {diag}}(\mathbf{s}_{i}) {\theta}_{i}\boldsymbol {\theta}
=\boldsymbol {\theta}^{\mathrm{H}}\tilde{\mathbf {B}}(\boldsymbol {\theta}\otimes\boldsymbol {\theta}),
\end{aligned}\end{equation}

\vspace{-3mm}
\begin{equation} \begin{aligned} 
\mathbf {v}^{\mathrm{H}} \mathbf {C}_{0} \mathbf {v}
=&\sum _{l=1}^{M}\boldsymbol {\theta}^{\mathrm{H}}\sum _{j=1}^{M} {\theta}_{j}^{\mathrm{H}}{\mathrm {diag}}(\mathbf{s}_{j}^{\mathrm{H}}){\mathbf {C}}_{0,j,l}{\mathrm {diag}}(\mathbf{s}_{l}) {\theta}_{l}\boldsymbol {\theta}
\\ =&(\boldsymbol {\theta}\otimes\boldsymbol {\theta})^{\mathrm{H}}\tilde{\mathbf{C}}(\boldsymbol {\theta}\otimes\boldsymbol {\theta}),
\end{aligned}\end{equation}
where $\mathbf{s}_{i}$ is the $i$th column of $\mathbf{S}$,
$\tilde{\mathbf {B}}=[\mathbf {B}_{0,1}{\mathrm {diag}}(\mathbf{s}_{1}), \ldots {},\\\mathbf {B}_{0,M}{\mathrm {diag}}(\mathbf{s}_{M})]$ and $\tilde{\mathbf {C}}=[\tilde{\mathbf {C}_{1}}{\mathrm {diag}}(\mathbf{s}_{1}),\ldots {},\tilde{\mathbf {C}_{M}}{\mathrm {diag}} (\mathbf{s}_{M})]$\\ with
$\tilde{\mathbf {C}_{l}}=[{\mathrm {diag}}(\mathbf{s}_{1}^{\mathrm{H}})\mathbf {C}_{0,1,l},\ldots {},{\mathrm {diag}}(\mathbf{s}_{M}^{\mathrm{H}})\mathbf {C}_{0,M,l}]^{\mathrm{T}}$. Based on the above definitions,
(P3) becomes 
\begin{equation}\begin{aligned}& \text {(P4)}\quad \underset { \boldsymbol {\theta}}{\min }~ \tilde{\boldsymbol {\theta}}^{\mathrm{H}}\tilde{ \mathbf {W}}\tilde{\boldsymbol {\theta}} 
    -2{\Re}(\boldsymbol {\theta }^{\mathrm{H}}\mathbf{p}^{\ast}+\boldsymbol {\theta}^{\mathrm{H}} \mathbf {D}_{0}^{\ast} \boldsymbol {\theta}^{\ast})
    \\&\qquad \quad ~\text {s.t.}~
     |\theta  _{i}|=1, \quad i\in \{1,\cdots,M\} ,
     \end{aligned}\end{equation}
     where  $\tilde{ \boldsymbol {\theta}}\!\!=\!\![\boldsymbol {\theta},\boldsymbol {\theta}\otimes\boldsymbol {\theta}]^{\mathrm{T}},\tilde{ \mathbf {W}}\!\!=\! \!\begin{bmatrix} \mathbf {A}_{0} &\tilde{ \mathbf {B}} \\ \tilde{ \mathbf {B}}^{\mathrm{H}} & \tilde{ \mathbf {C}} \end{bmatrix} $.
It can be readily found that $\tilde{ \mathbf {W}}$ is the Hermitian matrix. Then, in order to obtain the closed-form solution, a more tractable surrogate function can be constructed by the MM technique. 
Assuming the optimal $\boldsymbol{\theta}_{t}$ at the $t$th iteration, a surrogate function of $\tilde{ \boldsymbol {\theta}}^{\mathrm{H}} \tilde{ \mathbf {W}} \tilde { \boldsymbol {\theta}}$ w.r.t. $\tilde{\boldsymbol {\theta}}$
can be derived as \cite{7547360}
\begin{equation}\!\! \tilde{ \boldsymbol {\theta}}^{\mathrm{H}} \!\tilde{ \mathbf {W}}\! \tilde { \boldsymbol {\theta}}\!\leq \!\tilde{\boldsymbol {\theta}}^{\mathrm{H}}\!{\boldsymbol {\Lambda}}\tilde{\boldsymbol {\theta}}\!+\!2{\Re}\left(\tilde{\boldsymbol {\theta}}^{\mathrm{H}}\left(\tilde{ \mathbf {W}}\! -\!{\boldsymbol {\Lambda}}\right)\tilde{\boldsymbol {\theta}}_{t}\right)\!\!+\!\tilde{\boldsymbol {\theta}}_{t}^{\mathrm{H}}\left(\tilde{ \mathbf {W}} \!\!-\!\!{\boldsymbol {\Lambda}}\right){\tilde{\boldsymbol {\theta}}}_{t},\end{equation}
where $\boldsymbol {\Lambda}= {\lambda}_{ {max}}(\tilde{ \mathbf {W}}){\mathbf{I}}_{{M}^{2}+{M}}$. In order to avoid the high complexity induced by the eigenvalue decomposition of $\tilde{ \mathbf {W}}$, i.e., $\mathcal {O}({M}^{2}+{M})^{3}$, we choose $\boldsymbol {\Lambda}={\mathrm {tr}}(\tilde{ \mathbf {W}}){\mathbf{I}}_{{M}^{2}+{M}}={\mathrm {tr}}(\mathbf{A}_{0}){\mathbf{I}}_{{M}}+{\mathrm {tr}}(\tilde{\mathbf{C}}){\mathbf{I}}_{{M}^{2}}$ as another efficient auxiliary parameter. In addition, by employing the unit-modulus property of the RIS refection coefficients,  $\tilde{\boldsymbol {\theta}}^{H}{\boldsymbol {\Lambda}}\tilde{\boldsymbol {\theta}}$ can be transformed as ${\mathrm {tr}}(\tilde{ \mathbf {W}})({M}+{M}^{2})$, which is a constant term irrelevant to the optimization variable.  
 $\tilde{\boldsymbol {\theta}}_{t}^{\mathrm{H}}\left(\tilde{ \mathbf {W}} -{\boldsymbol {\Lambda}}\right){\tilde{\boldsymbol {\theta}}}_{t}$ is also a constant because ${\tilde{\boldsymbol {\theta}}}_{t}$ is already known at the ${t}$th iteration.
Therefore, the objective function in problem (P4) is simplified as 
\begin{equation}\begin{aligned}&
2{\Re}\left(\tilde{\boldsymbol {\theta}}^{\mathrm{H}}\left(\tilde{ \mathbf {W}} -{\boldsymbol {\Lambda}}\right)\tilde{\boldsymbol {\theta}}_{t}\right) -2{\Re}(\boldsymbol {\theta }^{\mathrm{H}}\mathbf{p}^{\ast}-\boldsymbol {\theta}^{\mathrm{H}} \mathbf {D}_{0}^{\ast} \boldsymbol {\theta}^{\ast})
\\=&2{\Re}(\boldsymbol {\theta}^{\mathrm{H}}\tilde{\mathbf {u}}_{t}+(\boldsymbol {\theta}^{\mathrm{H}}\otimes\boldsymbol {\theta}^{\mathrm{H}})\mathbf {V}_{t}-\boldsymbol {\theta}^{\mathrm{H}} \mathbf {D}_{0}^{\ast} \boldsymbol {\theta}^{\ast})
\\=&2{\Re}(\boldsymbol {\theta}^{\mathrm{H}}\tilde{\mathbf {u}}_{t}+\boldsymbol {\theta}^{\mathrm{H}}\tilde{\mathbf {V}}_{t}{\boldsymbol {\theta}^{\ast}}),
\end{aligned}\end{equation}
where $\tilde{\mathbf {u}}_{t}=(\mathbf {A}_{0}-{\lambda}_{ {max}}(\tilde{ \mathbf {W}}){\mathbf{I}}_{{M}})\tilde{\boldsymbol {\theta}}_{t}+\tilde{ \mathbf {B}}(\tilde{\boldsymbol {\theta}}_{t}\otimes \tilde{\boldsymbol {\theta}}_{t})-\mathbf{p}^{\ast},
    \mathbf {V}_{t}=\tilde{ \mathbf {B}}^{\mathrm{H}}\tilde{\boldsymbol {\theta}}_{t}+(\tilde{ \mathbf {C}} -{\lambda}_{ {max}}(\tilde{ \mathbf {W}}){\mathbf{I}}_{{M}^{2}})(\tilde{\boldsymbol {\theta}}_{t}\otimes \tilde{\boldsymbol {\theta}}_{t})$, and $\tilde{\mathbf {V}}_{t}=\hat{\mathbf {V}}_{t}-{\mathbf {D}_{0}}^{\ast}$.
    $\hat{\mathbf {V}_{t}}$ is the reshaped version of ${\mathbf {V}_{t}}$, i.e., ${\mathbf {V}_{t}}=\mathrm {vec}(\hat{\mathbf{V}_{t}})$.
Then problem (P4) is equivalently transformed into
\begin{equation}\begin{aligned}& \text {(P5)}\quad \underset { \boldsymbol {\theta}}{\min }~ 2{\Re}(\boldsymbol {\theta}^{\mathrm{H}}\tilde{\mathbf {u}}_{t}+\boldsymbol {\theta}^{\mathrm{H}}\tilde{\mathbf {V}}_{t}{\boldsymbol {\theta}^{\ast}})
    \\&\qquad \quad ~\text {s.t.}~
     |\theta  _{i}|=1, \quad i\in \{1,\cdots,M\}. 
     \end{aligned}\end{equation}
\vspace{-1mm}Using MM algorithm, we have transformed the original fourth-order problem (P4) into a second-order problem (P5). Such procedure greatly decreases the difficulty and complexity for solving the problem.
 However, problem (P5) is still non-convex. 
 To get the solution, we consider using MM algorithm again to find a more tractable surrogate function. To be specific, define 
$\overline{\boldsymbol{\theta }} \triangleq [\Re \lbrace \boldsymbol{\theta }^{T}\rbrace \, \, \Im \lbrace \boldsymbol{\theta }^{T}\rbrace ]^{T}$ and
    $\overline{\mathbf {V}}_{t} \triangleq \left[\begin{array}{cc}\Re \lbrace \tilde{\mathbf {V}}_{t}\rbrace &\Im \lbrace \tilde{\mathbf {V}}_{t}\rbrace 
    \\ \Im \lbrace \tilde{\mathbf {V}}_{t}\rbrace & -\Re \lbrace \tilde{\mathbf {V}}_{t}\rbrace \end{array}\right]$.  
The real function $\Re \lbrace \boldsymbol{\theta }^{\mathrm{H}}\tilde{\mathbf {V}}_{t}\boldsymbol{\theta}^{\ast}\rbrace$ can be expressed as $\overline{\boldsymbol{\theta }}^{\mathrm{T}}\overline{\mathbf {V}}\overline{\boldsymbol{\theta }}$.
Then, based on the second-order Taylor expansion, a convex surrogate function of $\overline{\boldsymbol{\theta }}^{\mathrm{T}}\overline{\mathbf {V}}\overline{\boldsymbol{\theta }}$ (i.e., $\Re \lbrace \boldsymbol{\theta }^{\mathrm{H}}\tilde{\mathbf {V}}_{t}\boldsymbol{\theta}^{\ast}\rbrace$) is derived as\cite{9769997}
\vspace{-2mm}  
\begin{equation}\begin{aligned} \overline{\boldsymbol{\theta }}^{\mathrm{T}}\overline{\mathbf {V}}\overline{\boldsymbol{\theta }} & 
   \!  \leq \!\overline{\boldsymbol{\theta }}_{t}^{\mathrm{\!T}}\overline{\mathbf {V}}_{t}\overline{\boldsymbol{\theta }}_{t}\! \!+\! \!\overline{\boldsymbol{\theta }}_{t}^\mathrm{T}(\overline{\mathbf {V}}_{\!t}\!\!+\!\!\overline{\mathbf {V}}_{\!t}^\mathrm{\!T}) (\overline{\boldsymbol{\theta }}\!\!-\!\!\overline{\boldsymbol{\theta }}_{t})
    \! \!+\! \!\frac{\lambda }{2}(\overline{\boldsymbol{\theta }}\!-\!\overline{\boldsymbol{\theta }}_{t})^\mathrm{\!T} \!(\overline{\boldsymbol{\theta }}\!-\!\overline{\boldsymbol{\theta }}_{t})
    \\ &=\Re \lbrace \boldsymbol{\theta }^\mathrm {H}\mathbf {U}\overline{\mathbf {v}}_{t}\rbrace +c, 
   \end{aligned} \end{equation}        
 where $\overline{\boldsymbol{\theta }}_{t} \triangleq [\Re \lbrace \boldsymbol{\theta }^\mathrm {T}_{t}\rbrace \, \, \Im \lbrace \boldsymbol{\theta }^\mathrm {T}_{t}\rbrace ]^\mathrm {T}$, $\lambda\triangleq{\lambda}_{ {max}}(\overline{\mathbf {V}}_{t}+\overline{\mathbf {V}}_{t}^\mathrm {T})$ and
  $\overline{\mathbf {v}}_{t}\triangleq (\overline{\mathbf {V}}_{t}+\overline{\mathbf {V}}_{t}^\mathrm {T}-\lambda \mathbf {I}_{2 M})\overline{\boldsymbol{\theta }}_{t} $.
$c$  is a constant irrelevant to $\boldsymbol{\theta }$.
 $\mathbf {U} \triangleq [\mathbf {I}_{M} \, \jmath \mathbf {I}_{M}]$ is defined to convert the real-valued function back to the original complex-valued function.
 So problem (P5) is equivalent to 
 \begin{equation}\begin{aligned}& \text {(P6)}\quad \underset { \boldsymbol {\theta}}{\min }~ 2{\Re}(\boldsymbol {\theta}^{\mathrm{H}}{\mathbf {f}_{t}}),
     \quad ~\text {s.t.}~
     |\theta  _{i}|=1, 
    \forall{i},
     \end{aligned}\end{equation} 
    where ${\mathbf {f}_{t}}=\tilde{\mathbf {u}}_{t}+\mathbf {U}\overline{\mathbf {v}}_{t}$. The optimal common RIS reflection coefficients are obtained as 
    \begin{equation} \boldsymbol {\theta}^{\mathrm{ opt}}=-e^{ j\text {arg}(\mathbf {f}_{t})}.
\label{24}        \end{equation}
In short, with the derivation of closed-form solutions for the transmit beamformers $\{ {\mathbf{F}}_{k} \}_{k=1}^{K}$, the receive equalizers $\{ {\mathbf{G}}_{k} \}_{k=1}^{K}$ and
the common RIS reflection pattern $\boldsymbol {\Theta }$, the overall optimization process is summarized in Algorithm 1.

 \renewcommand{\algorithmicrequire}{ \textbf{Input:}} 
\renewcommand{\algorithmicensure}{ \textbf{Output :}} 

    \begin{algorithm}[htb]
    \caption{ Proposed MM-based AO algorithm}
    \label{alg:example}
    \begin{algorithmic}[1] 
    \REQUIRE ~~$\mathbf {T}_{1}$,$\mathbf {T}_{2}$,$\mathbf {R}_{1,k}$,$\mathbf {R}_{2,k}$,$\mathbf {S}$, $P$,$\sigma _{n}^{2}$.
    \STATE {Initialize: ${\boldsymbol {\Theta} }^{0} = {\mathbf{I}_{M}}$,$ {\mathbf{F}}_{k}^{(0)} = [{\mathbf{I}}_{N_{k}}; {\mathbf{0}}_{(N_{t} - N_{k}) \times N_{k}}] $}.   
    \REPEAT 
    \STATE {Update $\lbrace{\boldsymbol {G}_{k}^{t+1} }\rbrace_{k=1}^{K}$  based on (\ref{6}).}
    \STATE {Update   $\lbrace{\boldsymbol {F}_{k}^{t+1} }\rbrace_{k=1}^{K}$  based on (\ref{9a}).}
    \STATE {Update $\boldsymbol{\theta}^{t+1}$ based on (\ref{24}).}
    \UNTIL {the convergence is satisfied.}
    \ENSURE ~~ $\lbrace{\boldsymbol {G} }_{k}^{\mathrm{ opt}}\rbrace_{k=1}^{K}$,$\lbrace{\boldsymbol {F} }_{k}^{\mathrm{ opt}}\rbrace_{k=1}^{K}$,${\boldsymbol {\Theta} }^{\mathrm{ opt}}$.  
    \end{algorithmic}
    \end{algorithm}

\vspace{-3.8mm}
    
    \subsection{Analysis of Convergence and Computational Complexity}
    \vspace{-0.6mm}
To analyze the convergence of the proposed MM-based AO algorithm, in the ${t}$th iteration, we first define the optimal solutions of its involved three subproblems  as $\lbrace{\boldsymbol {G} }_{k}^{t}\rbrace_{k=1}^{K}$, $\lbrace{\boldsymbol {F} }_{k}^{t}\rbrace_{k=1}^{K}$ and ${\boldsymbol {\Theta} }^{t}$ shown in (\ref{6}), (\ref{9a}) and (\ref{24}), respectively. The resultant objective value  of problem (P1) is given by $f(\lbrace{\boldsymbol {G} }_{k}^{t}\rbrace_{k=1}^{K}$, $\lbrace{\boldsymbol {F} }_{k}^{t}\rbrace_{k=1}^{K}$, ${\boldsymbol {\Theta} }^{t})$.  Then we have 
 \begin{equation}\begin{aligned}
 f(\!\lbrace{\boldsymbol {G} }_{k}^{t\!+\!1}\!\rbrace_{k\!=\!1}^{K},\lbrace{\boldsymbol {F} }_{k}^{t\!+\!1}\rbrace_{k\!=\!1}^{K},{\boldsymbol {\Theta} }^{t\!+\!1}) \!\!\leq  \!\!f(\lbrace{\boldsymbol {G} }_{k}^{t}\rbrace_{k\!=\!1}^{K},\lbrace{\boldsymbol {F} }_{k}^{t}\rbrace_{k\!=\!1}^{K},{\boldsymbol {\Theta} }^{t}),\label{23}
  \end{aligned}\end{equation}
which holds since the optimal closed-form solution of each subproblem in the ${t}$th iteration is available. It follows  from (\ref{23}) that the objective value of problem (P1) is monotonically non-increasing throughout the iterations.
In addition,  it is readily inferred that  the achievable  value of problem (P1)  is 
lower bounded by zero. As such, we can conclude that the proposed MM-based AO algorithm finally  converges to a locally optimal solution of problem (P1)\cite{Luo}.
Assuming the required number of iterations to be $I$, 
the total  complexity of the proposed algorithm is then calculated  as $\mathcal {O} \big (I ({K}( {N}_{k}^{3} +{M}^{3}+ N_{r,k}^{3}) +{8}{M}^{3}) \big)$.
\vspace{-3mm}
    \section{Numerical Result}
    \vspace{-1mm}
In this section, we demonstrate the simulations to show the performance of the MM-based AO algorithm. In a 3D Cartesian coordination, we assume the BS, RIS 1 and RIS 2 are deployed at (1,0,5)m, (0,0,2)m and (0,50,2)m, respectively. Two users   locate in a circle with the center at (1,50,0)m randomly. The numbers of BS transmit antennas  and each user receive antennas are 16 and 4, respectively. There are 64 reflecting elements in each RIS. Moreover, we assume 2 data streams are transmitted to each user. Unless otherwise specified, the noise power is set to be -120dBm and the BS maximum transmit power is ${P}=0$dBm. We assume Rician fading for each involved channel, i.e., $ {\mathbf{H}} = \sqrt { \beta } (\sqrt {\kappa } {\mathbf{H}}_{\mathrm{ LoS}} + \sqrt {1 - \kappa } {\mathbf{H}}_{\mathrm{ NLoS}}) $, where $\kappa $ is the Rician factor assumed to be 0.75 and $\beta$ is the distance-based path loss given by $\beta = \beta _{0} d^{-\gamma }$. $\beta _{0}$ denotes the reference path loss with ${d}=1$m and set as -30dB. For the BS $\rightarrow$ RIS ${1}$ and  RIS ${2\rightarrow}$ user ${k}$ links, the path loss exponents are given by ${\gamma }_{{T}_{1}}={\gamma }_{{R}_{2,k}}=2.2$. For other channels, the path loss exponents are given by ${\gamma }_{{T}_{2}}={\gamma }_{{R}_{1,k}}={\gamma }_{S}=3.6$.  ${\mathbf{H}}_{\mathrm{ NLoS}}$ is the small-scale fading component, obeying the Rayleigh distribution, i.e., ${\mathbf{H}}_{\mathrm{ NLoS}}\sim \mathcal {CN}(\mathbf{0},\mathbf{I})$. Moreover, our proposed algorithm is compared with the following benchmarks: (1) Single-RIS (BS side): There is only RIS 1 in the multi-user MIMO system. (2) Single-RIS (UE side): There is only RIS 2 in the multi-user MIMO system. (3) Double-RIS-Separate: The system model is the same with Fig. \ref{fig_1}, but the two RISs utilize two separate reflection coefficients.

First, the convergence of the MM-based AO algorithm is demonstrated in Fig. \ref{fig_2}. It's observed that the proposed algorithm converges within 15 iterations under different initializations, showing the superior convergence behavior.In addition, our proposed algorithm can achieve the same performance under different initialization schemes.
 \begin{figure}[htbp]
    \centering
    \includegraphics[width=2.5in]{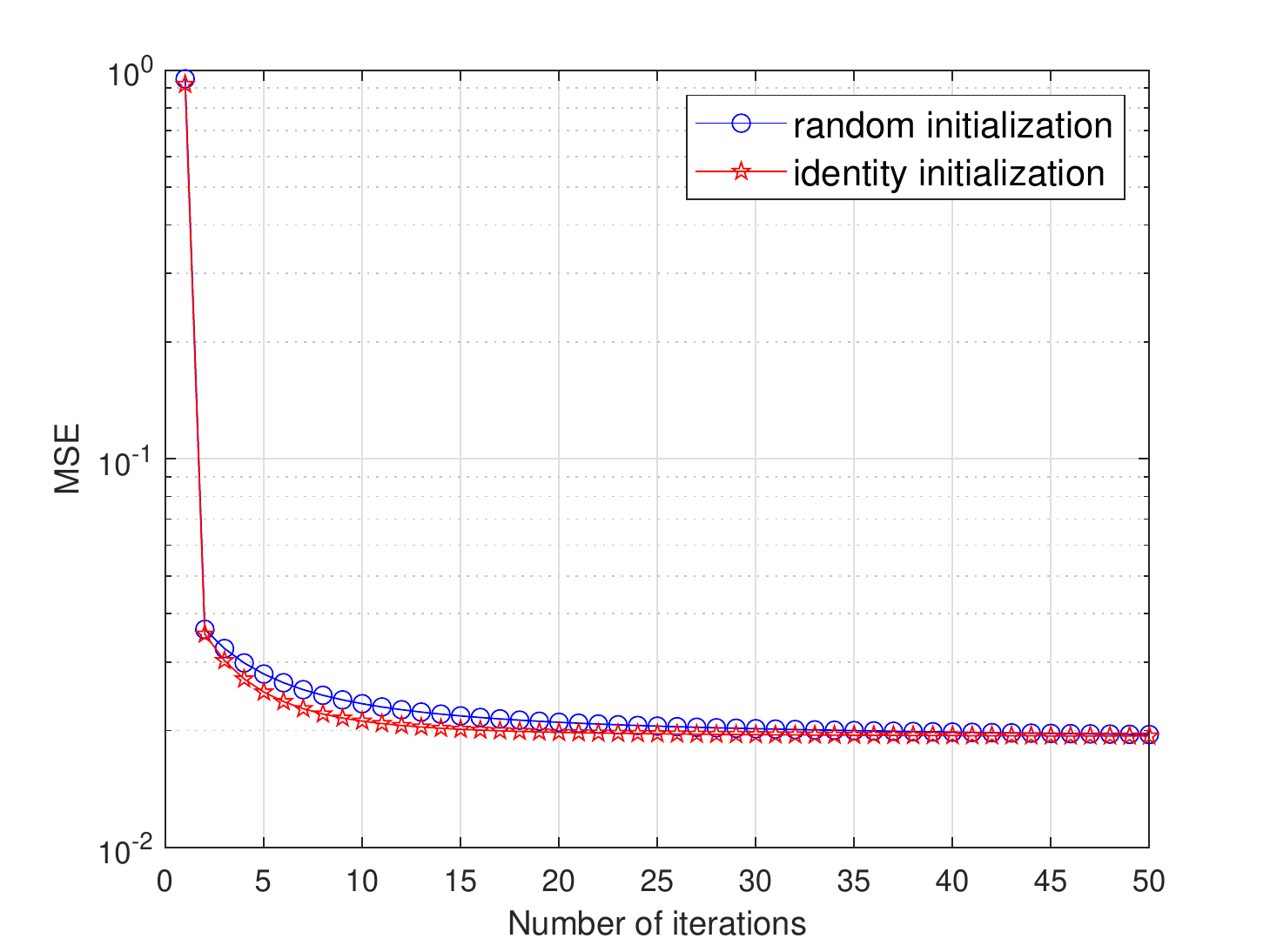}
    \vspace{-3mm}
    \caption{Convergence behavior of the proposed MM-based AO algorithm.}
    \vspace{-4mm}
   \label{fig_2}
    \end{figure} 
    \begin{figure}[htbp]
    \centering

    
    \includegraphics[width=2.5in]{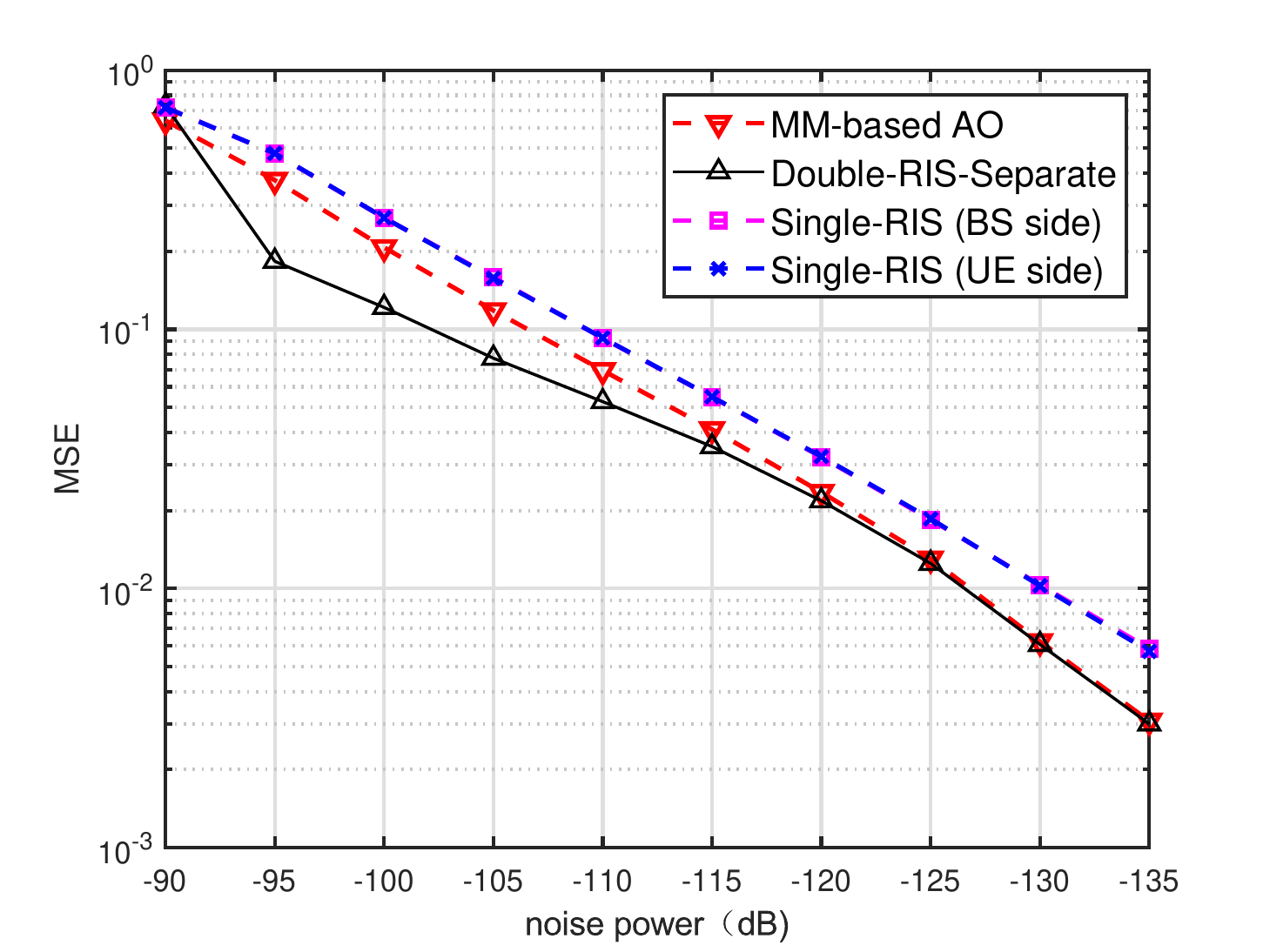}
    \vspace{-3mm}
    \caption{MSE performance versus
   the noise power for different schemes comparison.}
    \vspace{-4mm}
    \label{fig_3}
    \end{figure}
   
 Fig. \ref{fig_3} demonstrates the MSE performance of considered schemes versus the noise power.   
Compared with the Single-RIS schemes, i.e., Single-RIS (BS side) and Single-RIS (UE side), our proposed algorithm achieves lower MSE at the expense of the same signal processing  overhead. Furthermore, it is observed that our proposed algorithm can also achieve nearly 
the same performance as the  Double-RIS-Separate  scheme in the high-SNR region, even using  a half of signal processing overhead.
\vspace{-2mm}

 \begin{figure}[htbp]
    \centering
    \includegraphics[width=2.5in]{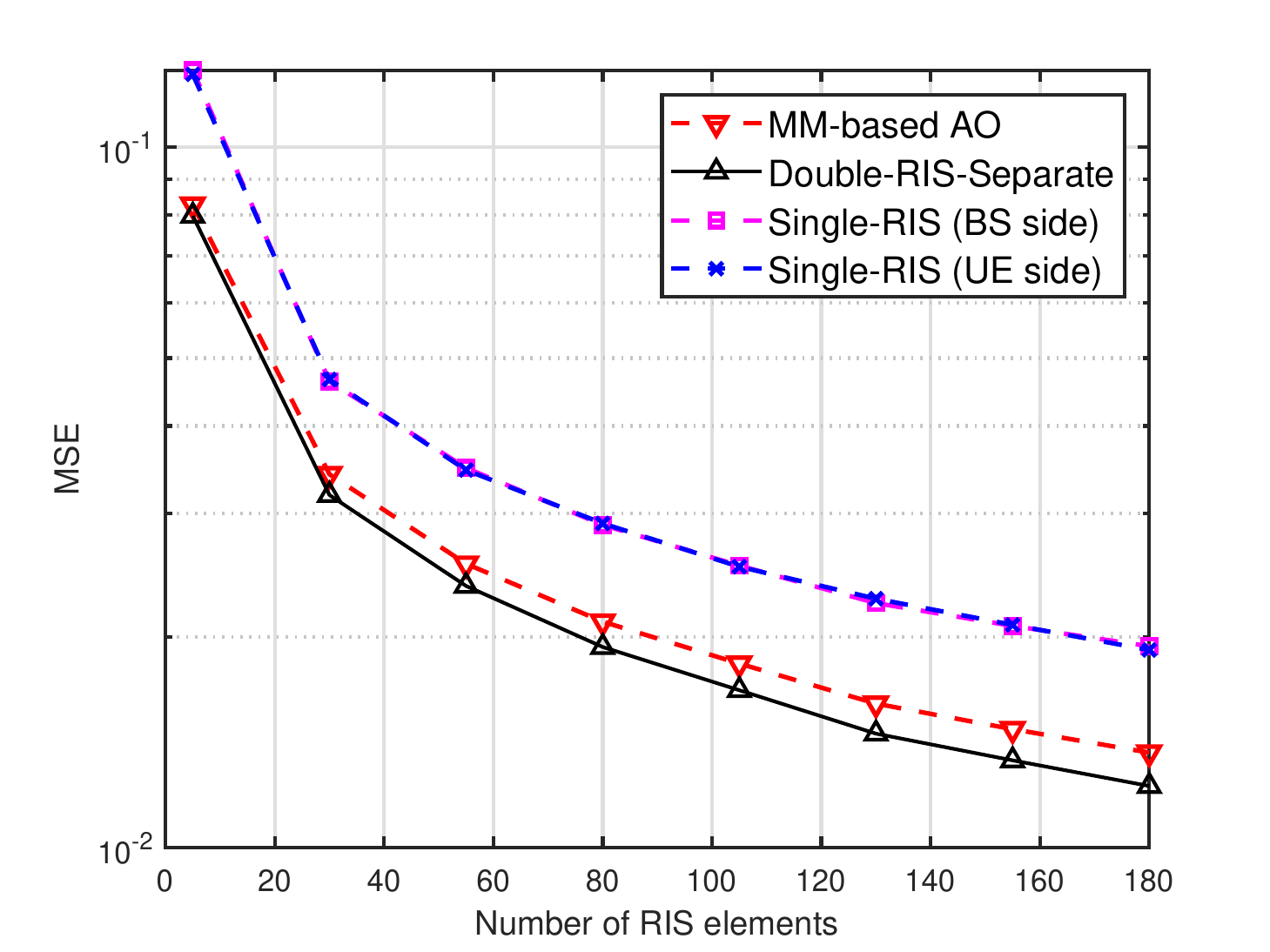}
   \vspace{-3mm}
    \caption{MSE performance versus
  the number of RIS  elements for different schemes comparison.}
  \vspace{-2mm}
    \label{fig_4}
    \end{figure}

In Fig. \ref{fig_4}, the MSE performance of  different schemes versus
  the number of RIS elements is demonstrated. All schemes exhibit better MSE performance as the number of RIS elements increases, which is attributed to the increased degrees of freedom. Observing from Fig. \ref{fig_4}, our system still outperforms the single-RIS systems and only has a slight performance loss as  compared to the double-RIS system  with separate reflection pattern. This indicates that the proposed  algorithm can strike a better tradeoff between the MSE performance and the implementation complexity of the considered system.
\vspace{-3mm}
 \section{Conclusion}    

This letter studied a cooperative double-RIS aided multi-user MIMO system, where the common RIS reflection pattern leading to the low communication overhead and signal processing complexity  is considered. We jointly optimized the active transmit beamforming at the BS, the equalizer at the receiver and the common reflection pattern at RISs by investigating the average sum MSE minimization problem. To solve this non-convex problem, we proposed an MM-based AO algorithm where the intractable original problem was decomposed into three subproblems. By exploiting the convex optimization theory and MM technology, each subproblem admitted a closed-form solution. Simulation results characterized the good convergence behavior of the proposed algorithm. Moreover, it  was shown  that the double-RIS with common reflection pattern achieved superior performance over the single-RIS systems and achieved nearly the same performance as the double-RIS with separate reflection pattern in the high-SNR region while the communication overhead is only half. This implies that a
superior tradeoff between the performance and complexity of the   implementation  and signal processing can be achieved for practical applications.
\vspace{-3mm}
\bibliographystyle{IEEEtran}
\bibliography{IEEEabrv,MSE}
\end{document}